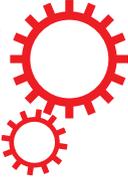

**OPEN**

# Ionic transport through sub-10 nm diameter hydrophobic high-aspect ratio nanopores: experiment, theory and simulation

Sébastien Balme[1,*], Fabien Picaud[2,*], Manoel Manghi[3,*], John Palmeri[4,*], Mikhael Bechelany[1], Simon Cabello-Aguilar[1], Adib Abou-Chaaya[1], Philippe Miele[1], Emmanuel Balanzat[5] & Jean Marc Janot[1]

Fundamental understanding of ionic transport at the nanoscale is essential for developing biosensors based on nanopore technology and new generation high-performance nanofiltration membranes for separation and purification applications. We study here ionic transport through single putatively neutral hydrophobic nanopores with high aspect ratio (of length $L = 6\,\mu$m with diameters ranging from 1 to 10 nm) and with a well controlled cylindrical geometry. We develop a detailed hybrid mesoscopic theoretical approach for the electrolyte conductivity inside nanopores, which considers explicitly ion advection by electro-osmotic flow and possible flow slip at the pore surface. By fitting the experimental conductance data we show that for nanopore diameters greater than 4 nm a constant weak surface charge density of about $10^{-2}$ C m$^{-2}$ needs to be incorporated in the model to account for conductance plateaus of a few pico-siemens at low salt concentrations. For tighter nanopores, our analysis leads to a higher surface charge density, which can be attributed to a modification of ion solvation structure close to the pore surface, as observed in the molecular dynamics simulations we performed.

Single solid-state nanopore technologies show great promise for bio-macromolecule detection or DNA sequencing[1]. For example, an important field of research is dedicated to developing nanopores that can uncover the changes in DNA that often play a role in cancer and other diseases and are hard to detect with current methods of sequencing. To carry out this program, different nanopore fields are currently being developed based on protein and synthetic nanopores[2]. Protein nanopores are certainly better suited for recognizing nucleotides (but suffer from liability), whereas synthetic nanopores can be manufactured on a large scale and be integrated into electronic devices. Synthetic nanopores still need to be better understood, however, before they can be perfected to the point of presenting the same advanced properties as protein ones[3–5]. The development of new generation high-performance nanofiltration membranes for important industrial applications, such as sea water desalination[6–8] is also being given a strong impetus from careful studies of transport through single nanopores.

[1]Institut Européen des Membranes, UMR5635 ENSCM-UM-CNRS, Place Eugène Bataillon, 34095 Montpellier cedex 5, France. [2]Laboratoire de Nanomédecine, Imagerie et Thérapeutiques, EA4662, Université Franche-Comté, Centre Hospitalier Universitaire, 16 route de Gray, 25030 Besançon cedex, France. [3]Université de Toulouse, Laboratoire de Physique Théorique (IRSAMC) UMR5152 CNRS-UPS, 118 route de Narbonne, F-31062 Toulouse, France,. [4]Laboratoire Charles Coulomb (L2C), UMR 5221 CNRS-Université de Montpellier, Montpellier, F-France. [5]Centre de recherche sur les Ions, les Matériaux et la Photonique, UMR6252 CEA-CNRS-ENSICAEN, 6 Boulevard du Maréchal Juin, 14050 Caen Cedex 4, France. *These authors contributed equally to this work. Correspondence and requests for materials should be addressed to S.B. (email: sebastien.balme@univ-montp2.fr)

| Nanopore | Length (μm) | Initial diameter (nm) | Bilayer number | ZnO cycles | Final diameter after HMDS grafting[a] (nm) | Experimental diameter[b] (nm) Phenom With correction | Experimental diameter[b] (nm) Phenom Without correction | Experimental diameter[c] (nm) No slip Without correction | Experimental diameter[d] (nm) Slip b=30 nm Without correction |
|---|---|---|---|---|---|---|---|---|---|
| $NP_1$ | 6 | 26.9 | 5 | 1 | 1 | 0.88 | 0.60 | 0.76 | 0.64 |
| $NP_{1.5}$ | 6 | 76.1 | 14 | 11 | 1.5 | 2.41 | 1.70 | 1.80 | 1.70 |
| $NP_2$ | 6 | 74.8 | 14 | 7 | 2 | 5.48 | 4.00 | 4.00 | 3.80 |
| $NP_5$ | 6 | 90.8 | 17 | 0 | 5.9 | 9.86 | 7.80 | 8.00 | 8.00 |
| $NP_{10}$ | 6 | 39.6 | 5 | 11 | 9.8 | 13.1 | 10.2 | 10.40 | 10.4 |

**Table 1.** Characteristics of the experimental nanopores and their measured diameters. [a]from initial diameter and thickness of the deposited layer,[b c d]from conductivity measurements: phenomenological with high concentration bulk correction, phenomenological without high concentration bulk correction no slip, model fit without high concentration bulk correction (without slip and with slip $b=30$ nm, see equations in the text).

Most studies focus on nanopores of the following types: (i) graphene[9,10], (ii) silicon nitride[11–13], (iii) polymeric track-etched[14,15] and (iv) carbon[16,17] or boron-nitride[18]. For all of them macromolecule detection is based on voltage clamp experiments where the recorded electrical signal is induced by ionic transport through nanopore. Understanding ionic transport at the nanoscale is thus of fundamental importance for developing and optimizing this technology. Indeed, ionic transport at the nanoscale is not like that in the bulk, since the conductance-concentration dependence does not follow a linear bulk-like behavior at low salt concentration. The best demonstration of this is the non linear law followed by the conductance reported for a nanoslit with a charged surface[19,20]. Two conductance regimes depending on salt concentration appear: (i) a constant conductance at low salt concentration, and (ii) a roughly linear increase at high salt concentration, similar to what appears in bulk like conditions. Other studies performed on single wall carbon nanotubes[16] (SWCNT), boron nitride nanotubes[18] (t-BNNT), PDMS-glass[21] and polymeric track-etched[22] nanopores have shown other interesting behaviors, such as stochastic ion pore blocking induced by nanoprecipitation, confined water, or nanopore surface wall wetting/dewetting.

Ionic transport through hydrophobic SWCNTs was investigated both experimentally[16,17] and via simulations[23]. This kind of nanopore, however, presents two major limitations: (i) a small diameter range, and (ii) the need for complicated experimental devices, which leads to non-negligible leakage currents[16]. The investigation of ionic transport through sub-10 nm hydrophobic nanopores needs an experimental methodology that requires: (i) designing small diameter nanopores with controlled surface states, and (ii) measuring the current without any leakage. To this end, we choose to combine track-etched nanopores with atomic layer deposition (ALD). Indeed, ALD is an outstanding technique for the deposition of conformal and homogenous ultrathin films due to its simplicity, reproducibility and the high homogeneity of the as-deposited films[24]. It allows the coating of flat surfaces and complex structures with a precise control of the thickness of the deposited film in the range of a few angstroms. Recently, we reported on the fabrication of sub-10 nm nanopores with long length (13 μm) using both track-etched techniques and ALD for tailoring hybrid biological solid state nanopores[5,25].

In the present work, we designed sub-10 nm hydrophobic nanopores and studied the ionic transport inside a single nanopore. We have built high aspect ratio hydrophobic nanopores of length $L=6$ μm with diameters ranging from 1 to 10 nm. We then measured their conductivity and showed, using molecular dynamics (MD) simulations and mesoscopic transport theory, that even for hydrophobic and organic surfaces, a small surface charge density is needed to interpret the experimental measurements.

## RESULTS

**Nanopore design.** Six nanopores (NPs) with high aspect ratios (length $L=6$ μm) were tailored by the single track-etched technique on PET film. The initial nanopore radii, $R$, were obtained from the dependence of the conductance, $G = \kappa \pi R^2/L$, on NaCl concentration, $c_s$, where $\kappa$ is the ionic conductivity, assuming bulk-like ionic conductivity ($\kappa = \kappa_b$) inside the nanopores at high salt concentration, i.e., assuming that $\kappa$ does not depend on the pore radius. Next, the number of $Al_2O_3$/ZnO bilayers (thickness equal to 2.48 nm)[25] and the number of ZnO cycles (thickness equal to 0.2 nm)[26] are adjusted to reach the expected final diameter (Table 1). The conformal coating of ALD on high aspect ratio pores has been first confirmed by Small-Angle X-ray Scattering (SAXS). A $q^{-4}$ slope is observed in the low $q$ region (Supporting information: figure SI-1)[25], characteristic of a sharp interface between the layers deposited by ALD and the air in pores. This slope confirms the quality of the deposition in terms of width control and homogeneity inside the pores. In order to approve the conformal coating, Transmission Electron Microscopy (TEM) was realized. The TEM image (Fig. 1) confirms the homogeneous ALD coating inside of a nanochannel of diameter 10 nm and of length 6 μm[27]. As shown by Cabello-Aguillar et al.[27], the nanopores can be considered homogenous along their length due to the homogeneous coating of ALD. The last step involved coating using hexamethyldisilazane (HMDS) to convert the hydrophilic surface due to the -OH terminal on ZnO to a hydrophobic Tri(Methyl-Silane) (TMS) function (($CH_3$)$_3$Si–) surface terminal group (Fig. 1). The surface chemistry of the nanopores has been studied by XPS and contact angle measurement. The ALD deposition results in a hydrophilic surface due to the -OH terminal on ZnO. By a hexamethyldisilazane (HMDS) treatment, this surface was converted to a hydrophobic (($CH_3$)$_3$Si–) surface terminal. The SAXS profile appears unchanged after surface conversion to (($CH_3$)$_3$Si–) surface terminal. Regarding the $q$ vector range accessible by our experimental set-up, it proves that the grafting did not alter the quality of the ALD coating. The success of the grafting has been attested to by XPS measurements (supporting information; table SI-1). Silicon in low content was detected (Si 2p binding energies 100.38 eV) that corresponds to a Si-$CH_3$ bond[25]. The hydrophobicity was confirmed by the determination of a contact angle of 92 °C on the treated HMDS surface[5,25].

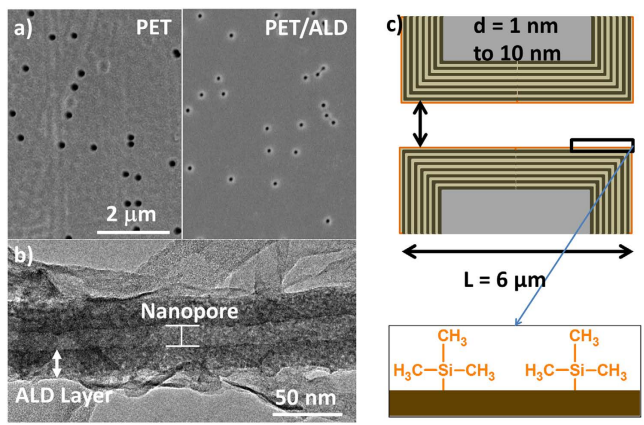

**Figure 1.** (**a**) SEM image of a multipore membrane before and after tuning by Atomic Layer Deposition (ALD), (**b**) TEM image of a single nanopore (10 nm of diameter and 6 μm in length) showing the conformal coating by ALD. The TEM image was obtained on multipore PET membranes (average pore diameter 70 nm) after 12 sequences of 5 cycles Al$_2$O$_3$/5 cycles ZnO deposited by ALD and followed by the elimination of the PET membrane, and (**c**) Schematic representation of a single nanopore design.

Therefore NP surfaces should be both uncharged and hydrophobic. In addition TMS plays the role of a passivation layer preventing the corrosion of ZnO[28] in NaCl medium. Indeed, without it, the nanopore gets blocked quickly during voltage clamp experiments.

**Influence of nanopore diameter on NaCl conductance.** The NaCl conductance through nanopores was studied by varying the NaCl concentration from $10^{-4}$ mol l$^{-1}$ to 5 mol l$^{-1}$. Experiments were performed using the current clamp method. Typically, the currents were recorded under a voltage range between 0 to 200 mV (Fig. 2a) with 10 mV step. Under these conditions the different nanopore sizes exhibit the linear current dependence on voltage predicted by Ohm's law (Fig. 2b,c,d). Note that the current trace exhibits metastable oscillations for NP$_2$ (only at high concentration) and for NP$_5$ (Fig. 2a). These oscillations can only be detected if the signal/noise ratio is high enough. Thus they likely exist for NP diameters smaller than 5 nm, but cannot be detected in high aspect ratio nanopores due to the resolution of our experimental setup. Experimentally, similar oscillations were previously reported in literature[21,29]. They could be interpreted as the oscillations between the stable and metastable branches of an ionic liquid-vapor phase transition[30–32].

For each NP, the conductance was determined as the slope of the linear dependence of the measured average current on applied voltage (Fig. 2). The results reported in Fig. 3 show that the NaCl conductance, $G(c_s)$, follows two regimes: (i) constant at low concentration, $c_s$ < 0.01 mol l$^{-1}$, and (ii) a roughly linear increase at high concentration, $c_s$ > 0.1 mol l$^{-1}$. This behavior is usually observed for charged nanopores[20]. Surprisingly, our experimental data show the same behavior, even if the TMS function at the pore surface is *a priori* not expected to exhibit charges.

For charged nanopores, the ionic transport inside the nanopore is usually described by the phenomenological equation: [20,21]

$$G = \frac{\pi R^2}{L}\kappa_b + G_r \qquad (1)$$

where $\kappa_b = e^2(\mu_+ + \mu_-)c_s$ is the bulk conductivity (uncorrected for ion-ion interactions), $e$ is the elementary charge, $\mu$ the ion mobility ($\mu_+$ for Na$^+$ and $\mu_-$ for Cl$^-$) and $G_r$ the (unknown) residual conductance. The bulk ionic mobilities $\mu_\pm = D_\pm/(k_B T)$ are related by the Einstein relation to the ionic diffusion coefficients, $D_+ = 1.334 \times 10^{-9}$ m$^2$ s$^{-1}$ (Na$^+$) and $D_- = 2.032 \times 10^{-9}$ m$^2$ s$^{-1}$ (Cl$^-$) (at 25°C) which are assumed to take on their dilute limit bulk values. In the bulk, ion-ion interactions play an increasingly important role at high salt concentration and therefore the bulk conductivity increases more slowly than linearly with increasing concentration. To account for this effect, which was neglected in previous work on nanopore conductivity, $\kappa_b$ should be multiplied by a correction factor $\chi(c_s) \leq 1$, which tends to 1 at low concentrations, and can be taken from experiment[33,34]. In small diameter nanopores, the long range hydrodynamic interactions that are thought to dominate this correction factor are strongly modified and probably less important than the ion-pore wall interactions. In such nanopores the correction factor will therefore depend on pore diameter, $D = 2R$, which is a topic for future investigation. For the time being we prefer to adopt a simplified approach and therefore compare the model predictions for the pore diameter obtained by fitting experimental conductance data using Equation (1) *with* and *without* the correction factor: although both choices allow us to account for the experimental data using $R$ and $G_r$ as fitting parameters, the model results for diameter without the correction factor appear to be in better agreement with those measured from the initial diameter and thickness of the deposited layer (see Table 1). We will therefore neglect, as is commonly done, the high concentration correction factor in our mesoscopic transport model presented below.

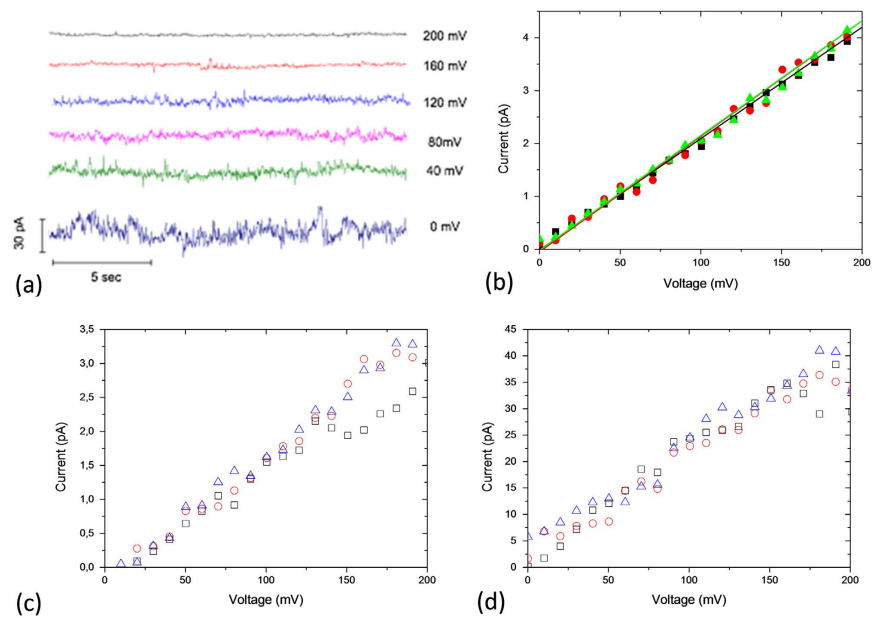

**Figure 2.** (**a**) Example of recorded current trace for $NP_5$ as a function of applied voltage for a NaCl concentration of 2.5 mol l$^{-1}$. Example of current-voltage (3 experiments) curve for (**b**) $NP_2$ at NaCl concentration 10$^{-3}$ mol l$^{-1}$, and $NP_5$ at NaCl concentration of (**c**) 10$^{-4}$ mol l$^{-1}$ and (**d**) 10$^{-2}$ mol l$^{-1}$.

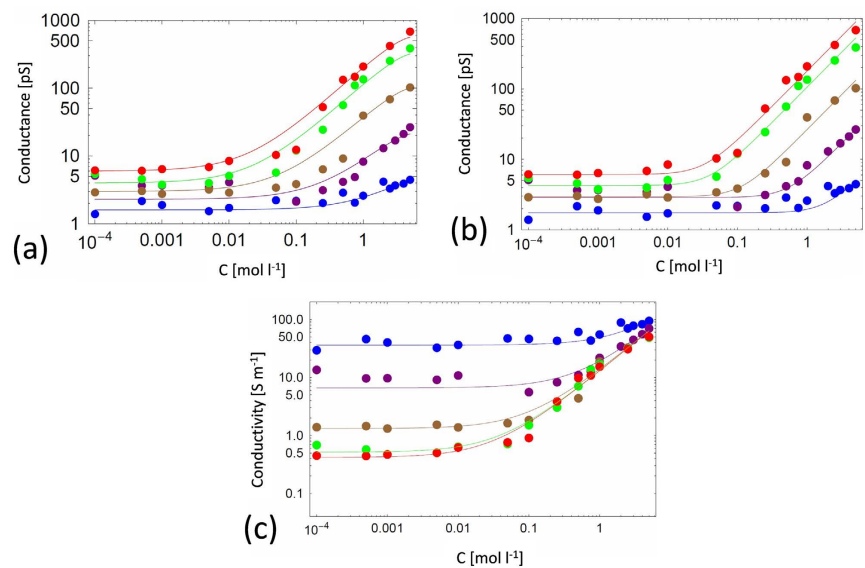

**Figure 3.** Nanopore conductance vs. NaCl concentration for, from bottom to top, $NP_1$, $NP_{1.5}$, $NP_2$, $NP_5$, $NP_{10}$: (**a**) phenomenological model fit with the bulk high concentration correction, (**b**) phenomenological fit without the bulk high concentration correction (**c**) nanopore conductivity vs. NaCl concentration without the bulk high concentration correction, from top to bottom, $NP_1$, $NP_{1.5}$, $NP_2$, $NP_5$, $NP_{10}$. Circles are the experimental data and solid lines are results from Eq. (1) (with the fitted diameters given in Table 1).

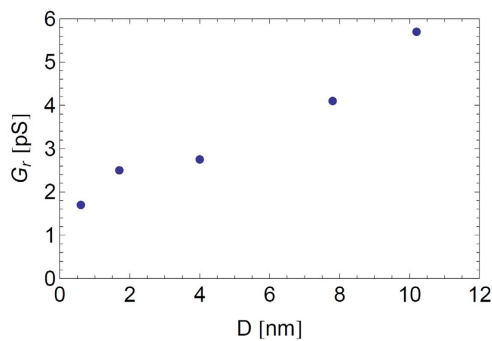

**Figure 4.** Phenomenological model fit values of the residual conductance, $G_r$, obtained from the fits of Fig. 3b at low salt concentration vs. nanopore diameter.

Limiting bulk ionic diffusivity (and therefore mobility) is determined by Stokes drag and dielectric solvent relaxation (of the ion induced local non-equilibrium water polarization) and there is evidence that approximately bulk values are found in nanopores with diameters greater than a critical value $D_c$ of roughly $3\,nm$[35–37]. The situation is yet not entirely clear, because evidence shows that the simulation boundary conditions may play an important role (periodic cylindrical nanopore geometry *vs* a nanopore connecting bulk-like reservoirs), but also that the nature of the nanopore surface may have a strong influence. Although it is often found that the effective ionic diffusion coefficient decreases with decreasing nanopore diameter (below the critical value), in one study of strongly hydrophilic nanopores with $D = 1\,nm$, the effective ionic diffusion coefficient along the pore axis has actually been found to exceed the bulk value[36]. Even when the ionic diffusion coefficients are found to decrease when decreasing the pore diameter below $D_c \approx 3\,nm$, this decrease is only about 25% down to $D = 0.7\,nm$[35,36]. There is also some evidence that if the pore diameter is corrected in a simple way to account for the actual nanopore water distribution, both the ionic diffusion coefficients and pore conductivity are not much different from their bulk values, even for diameters as small as $2\,nm$[38]. Faced with this situation, we assume here for simplicity that the ionic mobilities take on their dilute limit bulk values for all the nanopores studied here, which might lead to an overestimate (of not more than 25%) for the two tightest nanopores studied ($NP_1$ and $NP_{1.5}$).

The fits shown in Fig. 3 are reasonably good, with the expected linear bulk-like regime at high NaCl concentration range if the high concentration correction factor is neglected (Fig. 3a). At low concentrations the conductance attains a residual limiting plateau value, $G_r$, that varies roughly linearly with the nanopore diameter, as shown in Fig. 4. Hence, this residual conductance, which does not depend on $c_s$, allows us to account for the experimental data at low concentration.

The results reported in Table 1 show diameter values in qualitative agreement with the expected ones, based on the initial diameters and deposition thicknesses. The observed discrepancies, however, could be due to the rough evaluation of the initial diameter before coating. Indeed, the latter is determined assuming a bulk-like ionic transport.

Although necessary to fit the experimental data, this phenomenological residual conductance does not give any information concerning the origin of the low concentration conductance regime. We thus performed MD simulations of the system and developed a mesoscopic model to shed light on its microscopic origin.

**MD simulations of the experimental setup.** MD simulations have been shown to be a useful tool for gaining insight into transport of water and ions across nanopores[18,35–39]. In order to better understand our experimental observations, MD simulations were performed on neutral nanopores using several different diameters and NaCl concentrations, $c_s$. To study the structure of the different particles filling the nanopore, we calculated the radial concentration of each species during the MD simulations. This is defined as the number of particles present in a cylinder of radius $r$ and $r + \Delta r$ ($\Delta r = 0.5\,\text{Å}$) and of length $L$ over the volume of this cylinder. To avoid divergences at small radii, this concentration is weighted by the total concentration of each species in the pore. We plot in Fig. 5 the evolution of the concentration ratios for all the solution molecules (water and ions) as a function of the distance from the pore centre. Different initial concentrations and pore diameters were chosen to analyze carefully these distributions and to try to define a general behavior in this highly hydrophobic nanopore. From a general point of view, we show in Fig. 5 that the hydrophobicity of the neutral nanopore is clearly evident because of the total absence of water molecules over a distance larger than 0.12 nm from the last atom forming the nanopore. This distance was averaged over all the water molecules placed at the nearest distance from the atoms of neopentane molecules constituting the nanopore. The water nanopore diameters compared to the initial neopentane nanopore ones are summarized in Table 2. The effective radius of each nanopore could thus be rescaled since ions can only occupy the accessible volume offered by the water molecule location to enable the formation of a stabilizing water shell. The second general observation in Fig. 5 is the appearance of two water shells of different intensity. These two shells present the same amplitude for all the nanopore diameters indicating an identical behavior of the water molecules near the hydrophobic nanopore, as already observed in hydrophobic ionic channels such as KcsA[40]. (Smaller pore diameters were studied, but they lead to different regimes of pore filling, since we observed during the simulation a sudden absence of water molecules inside this highly confined system. The duration of this phase was too long to be considered here, and only a strong applied electric field could lead to a subsequent filling of the nanopore.)

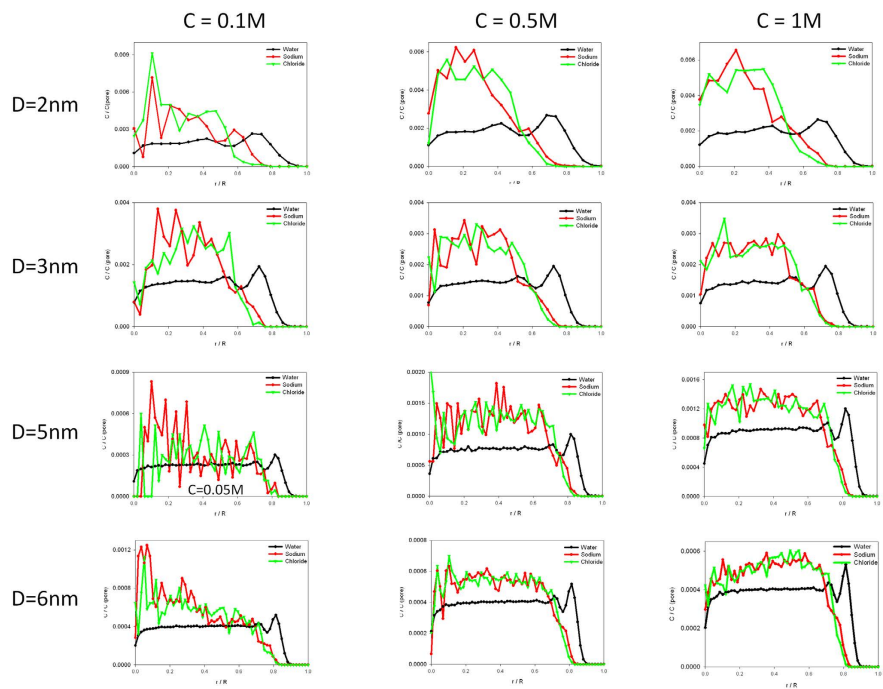

**Figure 5.** Concentration ratio profiles of ions and water in the nanopore, for various diameters and various concentrations, obtained by MD simulations. The concentrations are weighted by the total concentration of each species in the pore.

|  | NT2 | NT3 | NT5 | NT6 |
|---|---|---|---|---|
| Neopentane pore diameter (nm) | 1.92 | 2.64 | 4.63 | 5.37 |
| Water shell diameter (nm) | 1.80 | 2.48 | 4.40 | 5.21 |

**Table 2.** Water shell diameters as a function of the neopentane nanopore diameters.

With regard to the ion distributions, no difference between cations and anions could be uncovered as long as the pore diameter is large, $D > 2$ nm, or the ion concentration is higher than 0.5 mol l$^{-1}$. For $c_s < 0.5$ mol l$^{-1}$, or $D < 3$ nm, anion and cation shells can be differentiated, as shown in Fig. 5, where ions are placed more specifically in their water shell.

These numerical observations show quite complex ion/water profiles in the pore for small radii and thus bring new qualitative insights. In order to provide a physical and more quantitative interpretation of the experimental data, we can directly extract the surface charge density using an appropriate mesoscopic model for ion transport.

**Mesoscopic Poisson-Nernst-Planck model for conductivity.** We consider a monovalent salt (such as NaCl) in water solution at bulk concentration $c_s$, confined inside a cylindrical nanopore of radius $R$ and length $L$. The high aspect ratio of the experimental nanopores ($L >> R$), allows us to assume that the ionic concentrations in the pore are independent of the axial distance $z$ along the cylinder axis. Cations ($+$) and anions ($-$) concentrations inside the pore at distance $r$ from the center are given by

$$c_{\pm}(r) = c_s \, k_{\pm}(r) = c_s e^{\mp e\phi(r)/k_B T} , \qquad (2)$$

which defines the partition coefficients inside the pore, $k_{\pm}(r)$, where $\phi(r)$ is the electrical potential entering the radial Poisson-Boltzmann (PB) equation. Starting from the classical Poisson-Nernst-Planck (PNP) equations, the conductivity of the salt confined in the nanopore, $\kappa$, can be written as a function of the $k_{\pm}$:[41]

$$\kappa = \kappa^{em} + \kappa^{ad} = e^2 c_s \left( \mu_+ \overline{k}_+ + \mu_- \overline{k}_- \right) + \frac{(ec_s R)^2}{\eta} (\overline{k_+ g} - \overline{k_- g}) \qquad (3)$$

where $\eta = 8.94 \times 10^{-4}$ Pa s is the bulk water viscosity at 25 °C (it is assumed that both solvent viscosity and ionic mobilities in the nanopore are equal to their bulk values). The auxiliary function $g$ is

$$g(r) = \frac{1}{R^2} \int_r^R \frac{dr_1}{r_1} \int_0^{r_1} r_2 dr_2 [k_+(r_2) - k_-(r_2)] = \frac{2e}{k_B T} \left( \frac{\lambda_{DH}}{R} \right)^2 [\phi(r) - \phi(R)] \qquad (4)$$

and is written directly in terms of the PB electrical potential[41]. The Debye-Hückel length is $\lambda_{DH} = (8\pi l_B c_s)^{-1/2}$, where $l_B = \frac{e^2}{4\pi\epsilon k_B T} \simeq 0.7$ nm in water at room temperature is the Bjerrum length. The bar means that quantities are averaged in the pore, $\overline{y} = \frac{2}{R^2} \int_0^R y(r) r dr$.

The first term of Eq. (3), $\kappa^{em}$, is the electrical migration contribution associated with gradients in electrostatic chemical potential. In deriving Eq. (3) it is assumed, in accordance with the experimental set-up, that in the presence of an applied voltage difference, the concentration and pressure gradients across the nanopore vanish. These conditions lead to non-zero electro-osmotic solution flow through the nanopore, which is at the origin of the second term, $\kappa^{ad}$, i.e., the electro-osmotic contribution arising from the advection of ions induced by the flow inside the nanopore.

It is interesting to consider Eq. (3) in the homogeneous approximation, where $\phi(r)$ and therefore $k_{\pm}(r)$ are assumed to be constant in the pore. It is valid over the whole salt concentration range for $\sigma^* < 1$, where

$$\sigma^* = \pi l_B R \frac{\sigma}{e} \qquad (5)$$

is the dimensionless surface charge. In this limit, we have (see Appendix), $k_{\pm} = \overline{k}_{\pm} = e^{\mp e\phi_D/k_B T}$ where $\phi_D$ is the Donnan potential fixed by electroneutrality in the pore and created by the surface charge density $\sigma$ (taken to be negative): $\overline{k}_+ - \overline{k}_- = \frac{2|\sigma|}{eRc_s}$. The partition coefficients are thus

$$\overline{k}_{\pm} = \frac{|\sigma|}{eRc_s} \left[ \sqrt{1 + \left( \frac{eRc_s}{\sigma} \right)^2} \pm 1 \right]. \qquad (6)$$

The conductivity, Eq. (3), is then

$$\kappa_h = e^2 c_s (\mu_+ + \mu_-) \sqrt{1 + \left( \frac{\sigma}{eRc_s} \right)^2} + \frac{e|\sigma|}{R}(\mu_+ - \mu_-) + \frac{\sigma^2}{2\eta}. \qquad (7)$$

Two limiting cases are (i) the bulk case for $Rc_s \gg |\sigma|/e$, where $\overline{k}_{\pm} \approx 1$ and the 2 first terms (corresponding to $\kappa^{em}$) in Eq. (7) simplify to

$$\kappa_b = e^2 c_s (\mu_+ + \mu_-), \qquad (8)$$

which is the bulk conductivity of the salt; and (ii) the Good Co-ion Exclusion (GCE) limit, valid for $c_s \ll |\sigma|/(Re)$, which corresponds to the case where all co-ions are excluded from the pore, $\overline{k}_- \to 0$ and $\overline{k}_+ = 2|\sigma|/(eRc_s)$ (since $\sigma < 0$). The homogeneous GCE approximation for the conductivity is therefore

$$\kappa_h^{GCE} = \frac{2e|\sigma|}{R}\mu_+ + \frac{\sigma^2}{2\eta}. \tag{9}$$

which is independent of $c_s$. Although the first (electrical migration) contribution is exact within the scope of the PNP model at low salt concentration (because it depends only on global electro-neutrality in the pore), the second (advective) one is not: for sufficiently high $\sigma$ the homogeneous approximation breaks down because of strong radial variations in ion concentration (build up of counter-ions near a highly charged pore surface). To compute the advective part $\kappa^{ad}$ in the GCE limit valid over the whole range of $\sigma$, we solve the PB equation inside the pore within the GCE approximation[41] to obtain the PB potential $\phi_{GCE}(r)$:

$$k_+(r) = e^{-e\phi_{GCE}(r)/k_BT} = \left(\frac{\lambda_{DH}}{R}\right)^2 \frac{16\sigma^*(1+\sigma^*)}{\left(1+\sigma^*-\sigma^*\left(\frac{r}{R}\right)^2\right)^2}. \tag{10}$$

Thus $\overline{k}_+ = 16\sigma^*\left(\frac{\lambda_{DH}}{R}\right)^2 = \frac{2|\sigma|}{eRc_s}$, and, using Eq. (3) one obtains

$$\kappa_{GCE} = \frac{2e|\sigma|}{R}\mu_+ + \frac{\sigma^2}{2\eta}f(\sigma_*), \tag{11}$$

where the dimensionless charge density, $\sigma^*$, is defined in Eq. (5) and

$$f(\sigma^*) = \frac{2}{\sigma^*}\left[1 - \frac{\ln(1+\sigma^*)}{\sigma^*}\right]$$

is a monotonically decreasing function of $\sigma^*$ [$f(\sigma^* \to 0) \to 1$, $f(1) = 0.614$, and $f(\sigma^* \gg 1) \to 2/\sigma^*$] that accounts for corrections at large $\sigma^*$ to the advective part of the homogeneous GCE approximation, Eq. (9). As expected, $\kappa_{GCE}$ does not depend on $c_s$, but only on the surface charge density $\sigma$. For high $|\sigma|$ and/or large $R$ Eq. (11) simplifies to

$$\kappa_{GCE} \simeq \frac{2e|\sigma|}{R}\left(\mu_+ + \frac{1}{2\pi l_B \eta}\right) \tag{12}$$

Equation (12) is similar to the expression used in Eq. (1) of Ref. 14 to model the conductivity of KCl in highly charged boron nitride nanotubes at low salt concentration [although they used $\overline{\mu} = (\mu_+ + \mu_-)/2$ in place of $\mu_+$, a valid approximation for bulk KCl, because $\mu_{cl^-} \approx \mu_{K^+}$, but not for bulk NaCl, because $\mu_{cl^-} \approx 1.5\mu_{Na^+}$]. These two limiting behaviors, bulk-like Eq. (8) and GCE Eq. (9) [or Eq. (12)] are often simply added together in order to fit the experimental conductivity $\kappa(c_s)$, with or without the advective term in $1/\eta$, leading to an equation similar to Eq. (1).[18,42]

Experimental data for the conductance were fitted by adopting a hybrid approach, $G_{hyb} = \pi R^2 \kappa_{hyb}/L$, where the homogeneous approximation, Eq. (7), is used for the electrical migration part of the conductivity, and the GCE approximation, Eq. (11), for the advective one:

$$\kappa_{hyb} = e^2 c_s (\mu_+ + \mu_-) \sqrt{1 + \left(\frac{\sigma}{eRc_s}\right)^2} + \frac{e|\sigma|}{R}(\mu_+ - \mu_-) + \frac{\sigma^2}{2\eta}f(\sigma^*) \tag{13}$$

Within the PNP model this formula is exact in the homogeneous limit and in the full GCE regime (see Appendix) and should be an excellent approximation over the whole parameter range.

We analyzed the experimental conductivity by using Eq. (13), to extract the surface charge density $\sigma$ as a function of the pore radius $R$, which are both taken as fitting parameters (see Table 1 and Fig. 6a and b). We recall that the pore length was fixed at the experimental value of $L = 6\,\mu$m and the ionic mobilities and solvent viscosity were fixed at their bulk values.

For nanopore diameters greater than about 4 nm, the surface charge density is roughly constant with $\sigma \approx 0.02$ C m$^{-2}$ (as shown in Fig. 7, circles). This value is low with respect to the one estimated from experimental data for transmembrane boron nitride nanotubes[18] (t-BNNT).

This is not surprising, since the TMS function is supposedly not charged, contrary to the t-BNNT surface, which appears to exhibit an extremely high negative surface charge density at basic pH (~1 C m$^{-2}$ for pH > 10). The surface charge density of 0.02 C m$^{-2}$ obtained above is very close to the one reported by Shimizu[21] et al. (0.015 C m$^{-2}$) for native PDMS nanopores (sub-10 nm diameters). In view of the similar chemical structures for PDMS and TMS surfaces, similar values of surface charge density are coherent. This can be due to the adsorption of some OH$^-$ ions at the interface, as already observed at the water/air interface in Refs. 43–45. For nanopore diameters smaller than 4 nm, the surface charge density extracted from the model increases by about a factor of about 3 for $D = 0.76$ nm. One explanation for this behavior may come from the experimental nanopore design. Indeed, HDMS used for nanopore coating is performed under gas conditions; for small nanopore diameters, however, this coating might not be as homogeneous as it would be for larger nanopore diameters, due to a weaker permeability of the gas.

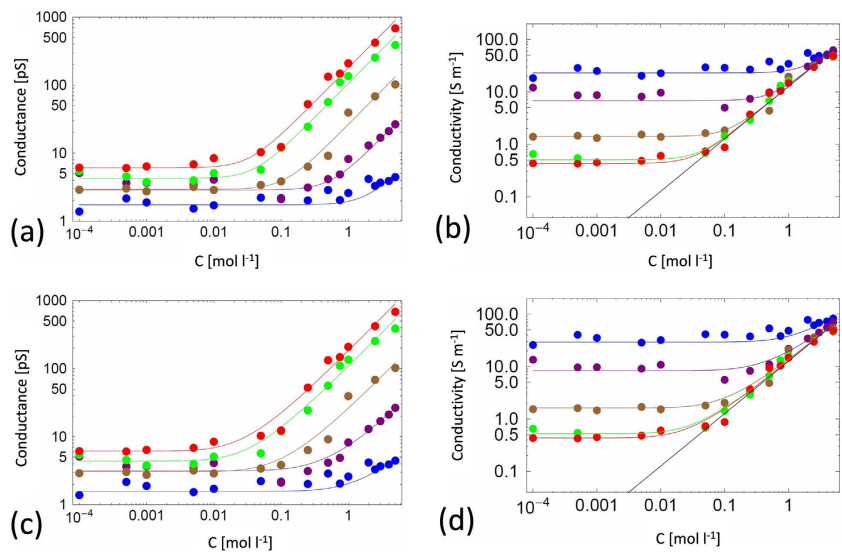

**Figure 6.** Experimental conductance data of Fig. 3 fitted using the hybrid mesoscopic approach: (**a**) conductance and (**b**) conductivity with slip length $b = 0$, using Eq. (13), (**c**) Conductance and (**d**) conductivity with slip length $b = 30$ nm, using Eqs. (13) and (14). The straight black line in figures (**b**) and (**d**) is the bulk conductivity (without the bulk high salt concentration correction factor, see Eq. 1 and below).

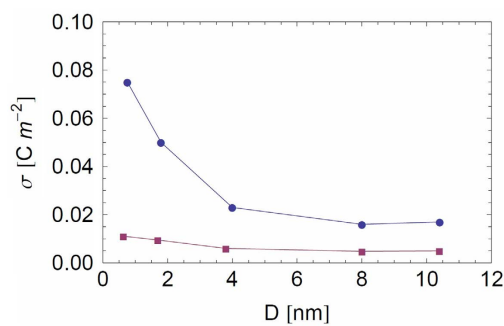

**Figure 7.** Surface charge of the nanopore as a function of nanopore diameter for $b = 0$ nm (circles) and $b = 30$ nm (squares): extracted from the mesoscopic model fits using Eqs. 13 and 14.

Since our nanopores are hydrophobic, it is interesting to consider the case where flow slip at the pore wall is taken into account, an effect previously studied experimentally and by molecular modeling under pressure driven flow for uncharged CNTs[46–48]. An important conclusion of the MD study is that for uncharged CNTs slip starts to play an increasingly important role when $R < 7$ nm[48]. Within the mesoscopic theory adopted here slip introduces a further contribution to the advective part, $\delta\kappa_{slip} = \frac{2\sigma^2 b}{\eta R}$, where $b$ is the *slip length*. Within the scope of the PNP model this result is exact, because it depends only on global electro-neutrality in the pore (see Appendix). In the homogeneous limit slip plays an extremely important role if $b > R/4$. For $b = 30$ nm, a typical value for *uncharged* hydrophobic nanopores,[49] slip may thus play a very important role for the nanopores under study, because $R \ll 4b \sim 120$ nm. Although MD simulation results indicate that electro-osmotic flows through sufficiently highly charged nanopores do not exhibit slip[37,39], the surface charge densities used in these studies were nearly an order of magnitude higher than those obtained above using our mesoscopic model without slip. The importance of slip for electro-osmotic flow in weakly charged nanopores is therefore a still open question. In order to examine the role of flow slip, we have thus analyzed the experimental conductance by adding the slip contribution to the advective one in Eq. (13),

$$\kappa_{slip} = \kappa_{hyb} + \delta\kappa_{slip} = \kappa_{hyb} + \frac{2\sigma^2 b}{\eta R},\quad(14)$$

with $b = 30$ nm to extract a modified nanopore surface charge density $\sigma$ as a function of $R$ (see Table 1 and Fig. 6c and d). We observe in Fig. 7 that the fitted surface charge densities are 3 to 6 times weaker for $b = 30$ nm than for the case with no slip and that the ratio between the highest and lowest values is reduced from 4 to 2 when slip is taken into account (for $D > 4$ nm, $\sigma \approx 0.005$ C m$^{-2}$). Because these trends make sense physically, we are tempted to conclude that for weakly charged nanopores introducing slip may be important in understanding the fitted surface charge densities from a microscopic point of view (e.g., adsorption of OH$^-$ ions at the pore surface). Note that the homogeneous approximation we used is reasonable since, without slip, the fitted values of $\sigma$ correspond to $0.4 < \sigma^* < 1.2$ for the nanopores NP$_1$ to NP$_{10}$, and $0.05 < \sigma^* < 0.4$ when slip is taken into account. These values are sufficiently low for the homogeneous approximation to be a reasonable one. The importance of the error incurred in the advective contribution without slip for the larger diameter nanopores (for which $\sigma^* = 0.09$ and $1.20$) is minimized because of the weaker role of advection compared with electrical migration ($< 30\%$) over the whole concentration range. In the case of slip (with $b = 30$ nm), the slip contribution to the advective term dominates at low salt concentration, where it is 2 to 6 times greater than the electrical migration one.

## DISCUSSION

The mesoscopic model developed above allows us to deduce the nanopore surface charge density, which varies with the nanopore radius for the smallest pore diameters. Two limitations of the model, however, should be discussed. Firstly, many nanopores are formed in membranes that have a dielectric constant $\varepsilon_m$, much smaller than that of bulk water ($\varepsilon_w \simeq 78$). A direct consequence is a dielectric exclusion of the ions inside a nanopore which increases with $\varepsilon_w/\varepsilon_m$. An important question concerns how this dielectric exclusion modifies the conductivity. Secondly, we neglected the fact that hydrophobic surfaces modify the water profile close to the nanopore surface. This also has an impact on the ionic profiles inside the nanopore, as seen in the MD simulations for neutral hydrophobic nanopores. It is expected, however, that pore wall surface charge will transform a neutral hydrophobic nanopore into an effective hydrophilic one, accompanied by further modifications of water and ion distribution and a reduction in flow slip[35–39,50]. It is also by now well established, and corroborated by our own MD simulations, that a strong enough applied electric field leads to water penetration of even strongly hydrophobic neutral nanopores that usually do not wet.

**Dielectric effects.** In order to improve our physical model of ion transport as applied to the actual experimental device, the role of the dielectric effects should be studied. Using the semi-homogenous approach of Ref. 32 where dielectric exclusion is studied using a variational method developed in Ref. 51, the averaged partition coefficients of the homogenous Donnan approach given in Eq. (6) are replaced by

$$\overline{k}_\pm = \frac{\sigma}{eRc_s}\left[\sqrt{1 + \left(\frac{eRc_s\Gamma}{\sigma}\right)^2} \pm 1\right]\quad(15)$$

where $\Gamma = \overline{\exp\left(-\frac{e^2}{2}w(r)\right)}$ is a coefficient that accounts for solvation and image corrections to the PB mean-field theory (or Donnan equilibrium), corresponding to $\Gamma = 1$. The quantity $e^2w(r)/2$ is the difference between the excess chemical potential of an ion located at a radial distance $r$ in the nanopore and the excess chemical potential of the same ion in the bulk: $w(r) = (\kappa_b - \kappa_v)l_B + \delta v_0(r, \kappa_v)$, where $\kappa_v$ and $\kappa_b$ are the inverse screening lengths in the pore and in the bulk, respectively; and $\delta v_0(r, \kappa_v)$ is the correction to the Debye-Hückel Green function due to the presence of the nanopore. It is important to note that $\Gamma$ does not depend on the surface charge density $\sigma$, is an increasing function of $c_s$ and $R$, and a decreasing one of the dielectric jump $\varepsilon_w/\varepsilon_m$.[43] Depending on the value of $c_s$, two limiting cases are of physical interest. For $R\Gamma c_s \ll \sigma/e$, which corresponds to a new dielectric GCE regime, we obtain small corrections to the electrical migration contribution to the conductivity $\kappa_h^{em}$

$$\kappa_{diel}^{em} = \kappa_h^{em} + \Gamma^2\frac{e^3Rc_s^2}{2\sigma}(\mu_+ + \mu_-) \approx \kappa_h^{em}\quad(16)$$

Hence for a charged nanopore, the limiting value of the electrical migration contribution to the conductivity for low $c_s$ is identical to the case without dielectric effects, Eq. (9). This comes about because this limiting value at low $c_s$ is determined uniquely by global electroneutrality and the counter-ion mobility inside the nanopore. However the advective part, $\kappa^{ad}$, will be modified by dielectric effects, although, for low $\sigma^*$, this contribution is weaker than $\kappa^{em}$, as explained above.

In the other limit, however, for which $R\Gamma c_s \gg \sigma/e$, the conductivity depends on the dielectric exclusion. Indeed, Eq. (15) simplifies to $\bar{k}_\pm = \Gamma \pm \frac{\sigma}{eRc_s}$, and the electrical migration contribution to the conductivity becomes

$$\kappa_{diel}^{em} = e^2 c_s \Gamma (\mu_+ + \mu_-) \simeq \Gamma \kappa_b, \qquad (17)$$

which is smaller than the mean-field value because $\Gamma \leq 1$. Indeed, due to dielectric exclusion, fewer ions enter the pore, which decreases the conductivity. A consequence of Eq. (17) is that for high $c_s$, the nanopore conductivity should follow $\kappa_{diel}^{em}$ and therefore slightly vary with the nanopore radius. This slight variation is not clearly seen in the data of Fig. 3, perhaps because of the noise in the data in the cross-over region between high-concentration bulk and low concentration plateau behavior. We note that dielectric exclusion for a *neutral* nanopore would lead to sub-bulk conductivity at low salt concentration, a result that is clearly incompatible with the observed plateau behavior.

**Ionic organization inside nanopore.** Theoretical developments show clearly that surface charge densities should be taken into account to interpret experiments and estimate the experimental diameter of the nanopores. The combined effects of flow slip length and dielectric exclusion tend to diminish the importance of the charges located on the pore surfaces, leading to very weak fitted surface charge densities that appear to be in better coherence with the actual experimental set up. While the slip length remains to be evaluated, the ionic concentration inside or outside the pore can be probed in MD simulations. Furthermore, at high bulk salt concentration the influence of the nanopore surface charge becomes negligible, and therefore it is interesting to study neutral nanopores. In Figure SI-2 we present the water and ionic profiles of Fig. 5 normalized by the bulk concentration as a function of $r/R$ for a neutral nanopore of radius $R = 3$ nm. As clearly observed, the water density profiles keep the same shape regardless of the ionic concentration inside the nanopore, whereas the ions density ones change.

The average concentrations in the pore change drastically compared to their bulk values, hence the curves shift. One can observe a dielectric exclusion of up to 40%, depending on $c_s$ and $R$. Taking into account dielectric exclusion in the nanopore conductivity at high salt concentration could then lead to modified fitted values for the nanopore diameters.

Moreover, a clear structuration of the water close to the surface and thus of the ionic profiles emerges for small pore radii (Figures SI-3). For large nanopore radii, the ionic distribution is almost homogenous, whereas it changes to a specific ionic surface one for strong confinement. This effect has obviously not been taken into account in our simplified continuum (mesoscopic) approach.

For charged nanopores, the object of future investigation, this phenomenon could play an extremely important role in the change in behavior of the measured conductivity from bulk-like to specific constant ones at low salt concentration.

## CONCLUSION

We have investigated ionic transport through hydrophobic nanopores with diameters in the sub 10 nm range. Our study brings together experimental, theoretical, and simulation approaches. The experimental data can be described using a simple phenomenological equation with a bulk-like conductivity at high salt concentration and a plateau at low. We developed a more sophisticated hybrid mesoscopic theoretical approach, in the Poisson-Nernst-Planck framework, for the calculation of the conductivity in nanopores and provided a simple fitting formula, Eq. (13), or Eq. (14) when flow slip at the pore wall is taken into account. By fitting the experimental data, a weak surface charge density has been determined and shown to remain constant for nanopore diameters greater than 3 nm. The higher surface charge density found for tighter nanopores could possibly be attributed to inhomogeneous hydrophobic coating. Molecular dynamics simulations indicate, however, a modification of ion solvation structure inside nanopores of less than 2 nm in diameter. Such a modification could also play a role in the formation of this higher surface charge density, although extensive MD simulations of charged nanopores will be needed to elucidate this phenomenon. Within a context of great enthusiasm and hope for nanopore technology, we hope that the present work will contribute to a better understanding of ionic transport at the nanoscale, which is of fundamental importance because it is at the origin of the electric signal recorded in applications where nanopores are used as sensors.

## METHODS

**Material.** NaCl was purchased from ACROS Organics (99,5%, 207790010). Ultra-pure water was produced from a Q-grad®-1 MilliQ system (Millipore). Poly(Ethylene Terephthalate) (PET) film (thickness 13 μm, biaxial orientation) was purchased from Goodfellow (ES301061). Diethyl Zinc (DEZ) ($Zn(CH_2CH_3)_2$, 95% purity, CAS: 557-20-0) and Trimethyl aluminum (TMA) ($Al(CH_3)_3$, 97% purity, CAS:75-24-1), were purchased from Sterm Chemical. Hexamethyldisilazane (HMDS) (reagent grade, ≥99%), was obtained from Sigma Aldrich

**Nanopore design.** *Track-etching of PET Film.* Single nanopores were tailored by track-etched method described elsewhere[5]. The single tracks were produced by $^{78}$Kr irradiation (8,98MeV) of PET film (6 μm) (GANIL, SME line, Caen, France). The etching protocol was performed as follows (i) UV exposition 24 h per side, (Fisher bioblock; VL215.MC, $\lambda = 312$ nm), (ii) chemical etching under NaOH solution (3 M, ~7 min, 50 °C) and (iii) 24h hours under ultrapure water.

*Nanopore reduction and surface functionalization.* $Al_2O_3$/ZnO ultrathin films were deposited using Atomic Layer Deposition (ALD). $Al_2O_3$ was obtained by alternating exposures of TMA and $H_2O$ at 60 °C with the following cycle times: 0.1 s pulse (TMA), 20 s exposure, and 40 s purge with dry Argon and a 2 s pulse ($H_2O$), 30 s exposure and 60 s purge. ZnO was fabricated using alternating exposures of DEZ and $H_2O$ with the following cycle times: 0.2 s pulse (DEZ), 20 s exposure, and 40 s purge with dry Argon and a 2 s pulse ($H_2O$), 30 s exposure and 60 s purge. The growth per cycle (GPC) was reported elsewhere to be 2 Å/cycle and 2.1 Å/cycle for $Al_2O_3$ and ZnO respectively[26,52]. Different sequence numbers of 5 cycles $Al_2O_3$ followed by 5 cycles ZnO, were applied to reduce the pore diameter from the initial to the final diameter. For $Al_2O_3$/ZnO nanolaminates an oxide bilayer thickness of 2.48 nm was determined by SAXS measurements inside the nanopore for each sequence, as reported elsewhere[5,25]. ZnO cycles were used on the top of $Al_2O_3$/ZnO nanolaminate in order to obtain the final desired diameters. After ALD deposition, all samples were functionalized by a 24 hours hexamethyldisilazane (HMDS) vapor exposure at room temperature in order to obtain hydrophobic surfaces. The expected result from the HMDS treatment was the replacement of the -OH bond on the surface of the ALD layer by a hydrophobic (($CH_3$)$_3$Si–) bond[19].

*Characterization.* Small-angle X-ray scattering, SAXS (Xenocs GenX equipped with a molybdenum anode and a MAR2300 2D imaging plate detector) has been used to determine the bilayer thickness. X-ray photoelectron spectrometry (XPS) measurement (ESCALAB 250 Thermo Electron) was used to approve the HMDS nanopores modification. TEM nanopore imaging was performed using a Transmission Electron Microscope (JEOL 2010). 6 μm thick multipore PET membrane (average pores diameter 70 nm) was characterized after 12 sequences of 5 cycles $Al_2O_3$/5 cycles ZnO deposited by ALD. The sample was annealed at 450 °C under air in order to remove the PET membrane.

**Ionic current measurements.** Ionic current measurements were performed using a patch-clamp amplifier (EPC800 LIH 8+8 and EPC10 HEKA electronics, Germany). The single nanopore was placed between two Teflon chambers containing NaCl solution from $10^{-4}$ M to 5 M. The current is measured by an Ag/AgCl electrode. Data were recorded at 10 kHz using Patchmaster software (Heka Elektronik, Germany). Recorded currents were analyzed by Fitmaster (Heka Elektronik, Germany).

**Molecular dynamics (MD) simulations.** All-atom molecular dynamics simulations were performed using NAMD 2.7b2 software[53]. To mimic experimental conditions, a constant temperature of 300 K (maintained by the Langevin dynamics method) and a constant pressure of 1 atm (maintained using the Langevin piston Nosé-Hoover method[54]) were imposed on the different simulated systems. The short- and the long-range forces were calculated every time step (1.0 fs). The classical particle mesh Ewald (PME) method[55] was used to evaluate the long-range electrostatic forces. The bond lengths between the hydrogens and the heavy atoms were constrained to their equilibrium values using the SHAKE/RATTLE algorithm[56,57].

CHARMM27[58] force field was used with TIP3P water model. The nanopore was mimicked by an assembly of neopentane molecules with fixed position for the central carbon atoms of neopentane, while the methyl groups constituting each molecule were free. Each neopentane was described according to Fig. SI-4. The total charge was set to zero but each atom possessed a local charge. The fixed central carbon atom has a charge equal to 0. Each carbon atom (hydrogen atoms respectively) belonging to methyl groups carries a -0.27e charges (0.09e respectively), according to Ref. 59.

The mean distance between the fixed centers of mass of neopentane molecules is chosen according to the work of Makowski *et al.*[60]. This nanopore was then inserted into a graphite bulk which was first cut to leave enough room for the pore. Several diameters (approximately equal to 2, 3, 5, and 6 nm) were studied and were denoted as NT2, NT3, NT5 and NT6, respectively. The whole system (containing the nanopore) was finally placed between two reservoirs and solvated at different ionic concentrations, ranging from 0.01 to 1 M, in order to study the ionic and water organization inside the nanopore, as described in Table SI 1.

All the simulation times were separated into an equilibration run of 15 ns and a production run of 15 ns to analyze the distribution of the ions in the different configurations. To increase the statistics when few ions were present in the simulations, 5 runs were simultaneously performed using different initial configurations.

## Acknowledgements

This work was supported in part by the French Research Program ANR-BLANC, Project TRANSION (ANR-2012-BS08-0023). Single tracks have been produced in GANIL (Caen, France) in the framework of an EMIR project. Molecular dynamic simulations were performed with the supercomputer regional facility Mesocenter of the University of Franche-Comté. We acknowledge R.R. Netz and L. Bocquet for enlightening discussions, Dr. J. Cambedouzou for SAXS measurements and Dr. M. C. Bechelany and Dr. F. Rossignol for TEM characterization.

## Author Contributions

S.B., S.CA. JM.J.: track-etched and patch-clamp experiments; A.AC., M.B., P.M.: experiment and characterization of Atomic layer deposition; E.B. : Single track experiment. F.P: molecular dynamic simulations; M.M, J.P.: theoretical physics; S.B., F.P., M.B., M.M., J.P.: Design study, data interpretation, wrote the paper.

## Additional Information

**Supplementary information** accompanies this paper at http://www.nature.com/srep

**Competing financial interests:** The authors declare no competing financial interests.

**How to cite this article**: Balme, S. *et al*. Ionic transport through sub-10 nm diameter hydrophobic high-aspect ratio nanopores: experiment, theory and simulation. *Sci. Rep.* **5**, 10135; doi: 10.1038/srep10135 (2015).